\documentclass[aps, pre, twocolumn, showpacs, superscriptaddress, groupedaddress, amssymb]{revtex4-1}
\usepackage{graphicx}
\usepackage{amsfonts}
\usepackage{amssymb}
\usepackage{float}
\usepackage{latexsym}
\usepackage{amsmath}

\usepackage{subfigure}

\begin{document}

\title{Dust cluster spin in complex (dusty) plasmas}

\author{Bo Zhang}
\affiliation{Center for Astrophysics, Space Physics, and Engineering Research (CASPER), Baylor University, Waco, Texas 76798-7310, USA}
\author{Jie Kong}
\affiliation{Center for Astrophysics, Space Physics, and Engineering Research (CASPER), Baylor University, Waco, Texas 76798-7310, USA}
\author{Mudi Chen}
\affiliation{Center for Astrophysics, Space Physics, and Engineering Research (CASPER), Baylor University, Waco, Texas 76798-7310, USA}
\author{Ke Qiao}
\affiliation{Center for Astrophysics, Space Physics, and Engineering Research (CASPER), Baylor University, Waco, Texas 76798-7310, USA}
\author{Lorin S. Matthews}
\affiliation{Center for Astrophysics, Space Physics, and Engineering Research (CASPER), Baylor University, Waco, Texas 76798-7310, USA}
\author{Truell W. Hyde}
\affiliation{Center for Astrophysics, Space Physics, and Engineering Research (CASPER), Baylor University, Waco, Texas 76798-7310, USA}

\date{\today}

\begin{abstract}
The spontaneous rotation of small dust clusters confined inside a cubical glass box in the sheath of a complex plasma was observed in experiment. Due to strong coupling between the dust particles, these clusters behave like a rigid-body where cluster rotation is contingent upon their configuration and symmetry. By evaluating the effects of distinct contributing forces, it is postulated that the rotation observed is driven by the net torque exerted on the cluster by the ion wake force. The configuration and symmetry of a cluster determines whether the net torque induced by the ion wake force is nonzero, in turn leading to cluster rotation. A COPTIC (Cartesian mesh, oblique boundary, particles and thermals in cell) simulation is employed to obtain the ion wake potential providing a theoretical model of cluster rotation which includes both the ion wake force and neutral drag and predicts rotation rates and direction in agreement with experimental results. These results are then used to diagnose the ion flow within the box. 
\end{abstract}

\maketitle

\section{Introduction}
Complex plasmas consist of small solid microparticles immersed in a plasma environment, and are the subject of widespread interest across a rich variety of research fields ~\cite{fortovbook, chu94, thomas96, fortov04, shuklaRev, baumgartner09, bonitz10, hartmann10, hartmann14}. Once injected into the plasma, microparticles become negatively charged due to the greater thermal velocity of the electrons compared to the ions. The particles interact through a shielded Coulomb potential, and many different dust structures in a plasma have been realized in ground-based laboratories and under microgravity conditions on board the International Space Station (ISS). Examples of these structures include vertical strings ~\cite{melzer06, kong11}, Yukawa or Coulomb balls ~\cite{arp04, bonitz06}, 2D ~\cite{knapek07, nosenko09} and quasi-2D systems ~\cite{woon04, chan07} and 3D ~\cite{klumov10, khrapak11} Coulomb crystals. Among the various strongly-coupled systems formed by dust particles, multiple vertical strings were recently utilized as a system for diagnosing structural phase transitions by Hyde \textit{et al.} ~\cite {kong13}. 

Dusty plasma systems display interesting collective effects such as vortices ~\cite{vaulina03, schwabe14}, dust acoustic waves (DAW) ~\cite{schwabe07, menzel10}, and dust rotation ~\cite{yousefi14, nosenko, worner11, worner12, laut, konopka00, sato01, cheung04, hou05, huang11, carstensen09, kahlert12, hartmann13, schablinski14}. Rotational dust motion is generally classified into one of two categories, rigid-body rotation $\Omega(\rho) \simeq $ const, with $\rho$ the rotation radius~\cite{nosenko, worner11, worner12, laut,  sato01, cheung03, cheung04, hou05, huang11, carstensen09, kahlert12, hartmann13}, and sheared differential rotation ~\cite{konopka00, schablinski14, klindworth00}. 

In the majority of cases presented in the literature to date, cluster rotation has been shown to be driven by externally controlled parameters triggered by rotating electric fields ~\cite{nosenko, worner11, worner12, laut}, axial magnetic fields ~\cite{konopka00, sato01, cheung03, cheung04, hou05, huang11}, or rotating electrodes ~\cite{carstensen09, kahlert12, hartmann13, schablinski14}. Cheung \textit{et al}. applied an axial magnetic field to induce dust cluster rigid-body rotation, proposing that the radial confinement electric field is modified by the magnetic field, which in turn changes the angular velocity of the dust cluster ~\cite{cheung03}. Klindworth \textit{et al}. found that structural transitions, together with intershell rotation of the cluster, can be excited by exerting a torque on the cluster using two opposing laser beams. They also found that the decoupling of the shells within these finite clusters can occur creating a transition from cluster to intershell rotation by altering the Debye shielding. In this case, the intershell rotation barrier of a sixfold cluster is about twice as large as the Coulomb case ~\cite{klindworth00}. 

Recently, two innovative techniques using rotating electrodes and rotating electric fields were employed to investigate dust cluster rotation. The first technique is based on the assumption that the effects of the Coriolis force $2m(\vec{v} \times \vec{\Omega})$ and the Lorentz force $Q(\vec{v} \times \vec{B})$ are equivalent, allowing the study of magnetic field effects on complex plasmas without the necessity of installing a high power magnet setup ~\cite{carstensen09, kahlert12, hartmann13, schablinski14}. This technique allows experiments to be implemented through adoption of a rotating electrode to set the background neutral gas into rotation, with the subsequent gas drag driving the dust cluster rotation ~\cite{schablinski14}. Another interesting technique for driving clusters into rotation is through the use of a rotating electric field (see Refs. ~\cite{nosenko, worner11, worner12, laut}). Rotation is sustained by combining the torque created by the ion-drag and the field generated by the rotating electric field ~\cite{worner11}. 

In the present paper, structures consisting of multiple vertical strings are used as probes to study the spontaneous rotation of clusters trapped in a glass box. Rotations of clusters having asymmetric configurations are observed to take place naturally. We propose that such spontaneous cluster rotation is caused by the torque due to the ion wake force exerted on the asymmetric cluster. This allows the ion flow to be investigated using the rotation of the cluster. 

\section{Experimental setup}
The experiment described here was conducted in a modified gaseous electronics conference (GEC) radiofrequency (rf) discharge cell ~\cite{kong14, qiao14}. The lower electrode has a diameter of 8 cm and is capacitively coupled to a rf signal generator operated at a frequency of 13.56 MHz. The upper electrode consists of a ring having a diameter of 8 cm, which is grounded, as are the surrounding cell walls. The vertical separation between the upper and lower electrodes is 1.9 cm. A dust dispenser above the grounded ring serves to introduce dust particles into the plasma, with oscilloscopes used to monitor the rf voltage and self-bias generated at the lower rf electrode. All experiments were conducted in argon gas at pressures between 100 and 200 mTorr. 

\begin{figure}[tbp]
\includegraphics [width=0.48\textwidth]{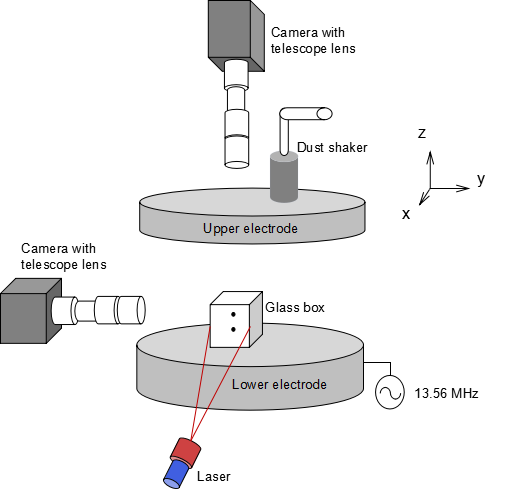}
\caption{Sketch of the experimental setup. The plasma discharge is operated between the grounded ring-shaped upper electrode and the lower rf electrode which is driven at 13.56 MHz.}
\label{fig:ITObox}
\end{figure}

Melamine formaldehyde (MF) microparticles having a mass density of 1.514 g/cm$^3$ and a diameter of 8.89 $\mu$m (as supplied by the manufacturer) were used. Particles were illuminated employing either a vertical or horizontal sheet of laser light. A Sony XC-HR50 charge-coupled device (CCD) camera operated at a frame rate of 60 fps and a Photron Fastcam 1024 PCI high-speed camera operated at a frame rate of 250 or 500 fps, were used to record the trajectories of the dust particles. 

In all experiments, the dust particles were confined in an open-ended glass box with a height of 12.7 mm and a width of 10.5 mm placed on the powered lower electrode ~\cite{kong14}, as shown in Fig. \ref{fig:ITObox}. 

\begin{figure}[tbp]
\begin{center}
\includegraphics [width=0.48\textwidth]{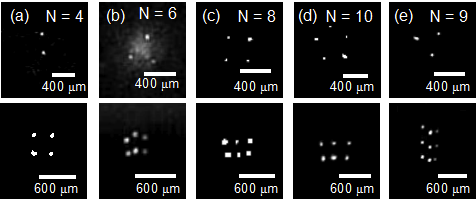}
\caption{Top (upper row) and corresponding side (lower row) views of symmetric dust structures formed inside a glass box placed on the lower electrode. The system parameters and number of particles for each configuration are (a) 2.04 W, 100 mTorr, $N = 4$, (b) 2.17 W, 150 mTorr, $N = 6$, (c) 2.56 W, 150 mTorr, $N = 8$, (d) 2.19 W, 150 mTorr, $N = 10$, and (e) 2.43 W, 150 mTorr, $N = 9$.}
\label{fig:symmetricStruc}
\end{center}
\end{figure}
  
 \section{Results}
Multiple string structures were observed to form inside the glass box for neutral gas pressures between 100 and 200  mTorr and rf powers between 1.37 and 5.92 W. Dust cluster symmetry was observed to determine spontaneous rotation, with this rotation directly related to dust particle configuration. Symmetric cluster configurations were observed to exhibit little or no rotation; however, when this symmetry was broken, spontaneous rotation of the cluster was observed. 

Cluster symmetry was determined primarily by the number of particles and system confinement. In this case, symmetric structures were formed using a glass box of cubical geometry, which provides an isotropically harmonic trap potential in the central region of the box ~\cite{nosenko, worner11, worner12, laut}. Fig. \ref{fig:symmetricStruc} (a)-(e) shows a series of representative symmetric structures formed in this manner with (a)-(d) showing symmetric multiple-string structures consisting of dust particles arranged as two to five, two-particle strings. A three, three-particle chain structure comprised of nine particles is presented in Fig. \ref{fig:symmetricStruc}  (e). No appreciable rotation for any of the clusters shown in Fig. \ref{fig:symmetricStruc} was observed.  

\begin{figure*}[tbp]
\begin{center}
\includegraphics [width=0.90\textwidth]{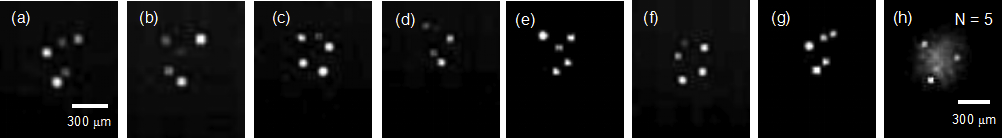}
\caption{Sequence of images illustrating the rotation of an asymmetric structure with $N = 5$ particles. (a)-(g) correspond to the side view of this system with $\Delta t = 0.3$ s between frames. The top view of the cluster is shown in (h). System parameters are 1.76 W, 150 mTorr.}
\label{fig:asymmetricStruc}
\end{center}
\end{figure*}

\begin{figure}[tbp]
\begin{center}
\subfigure{
\includegraphics [width=0.16\textwidth]{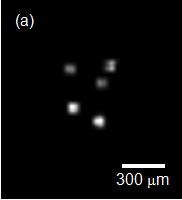}
}
\hspace{29.8pt}
\subfigure{
\includegraphics [width=0.16\textwidth]{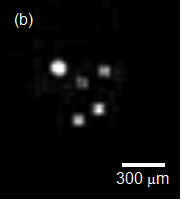}
}
\subfigure{
\includegraphics [width=0.22\textwidth]{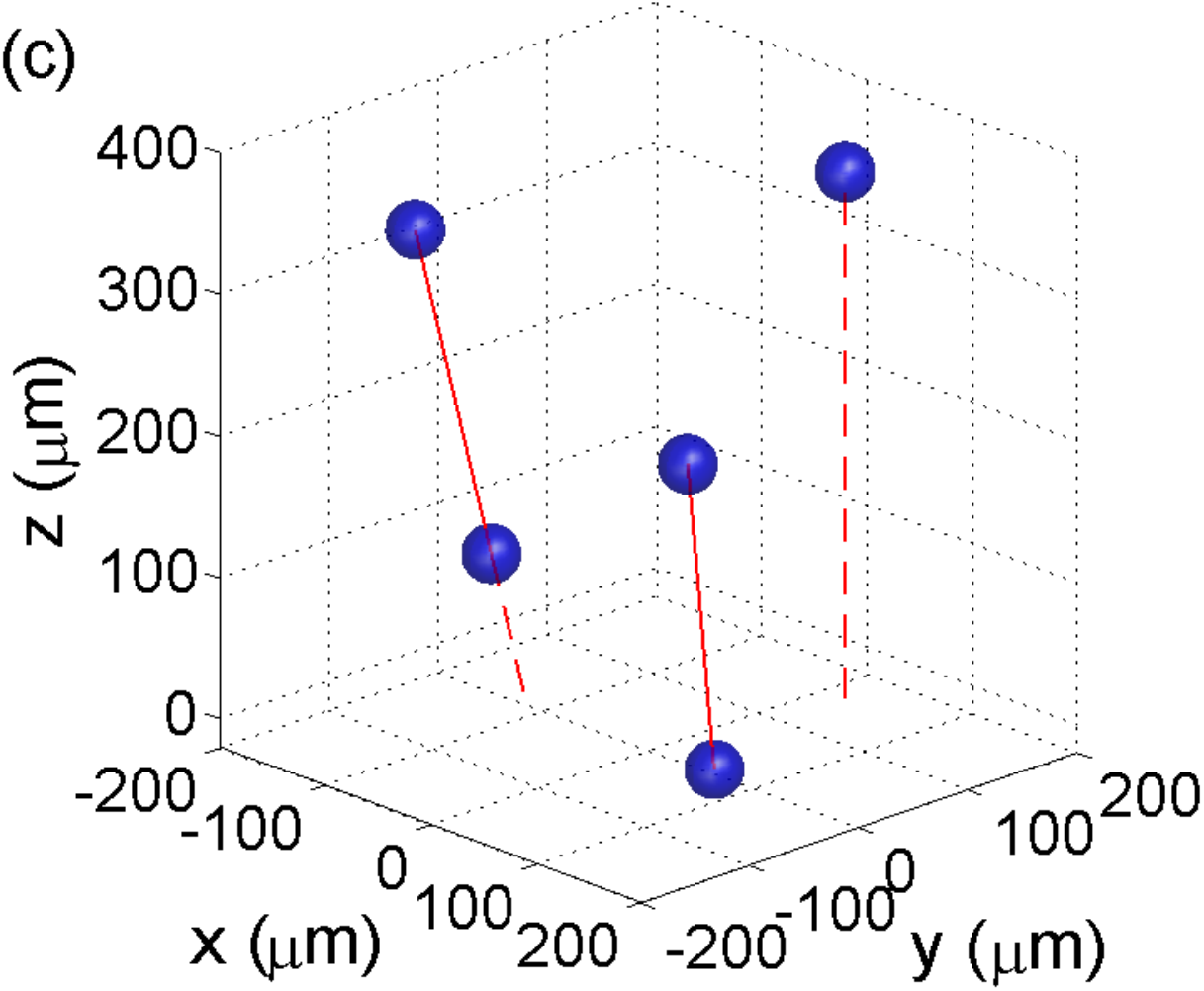}
}
\subfigure{
\includegraphics [width=0.22\textwidth]{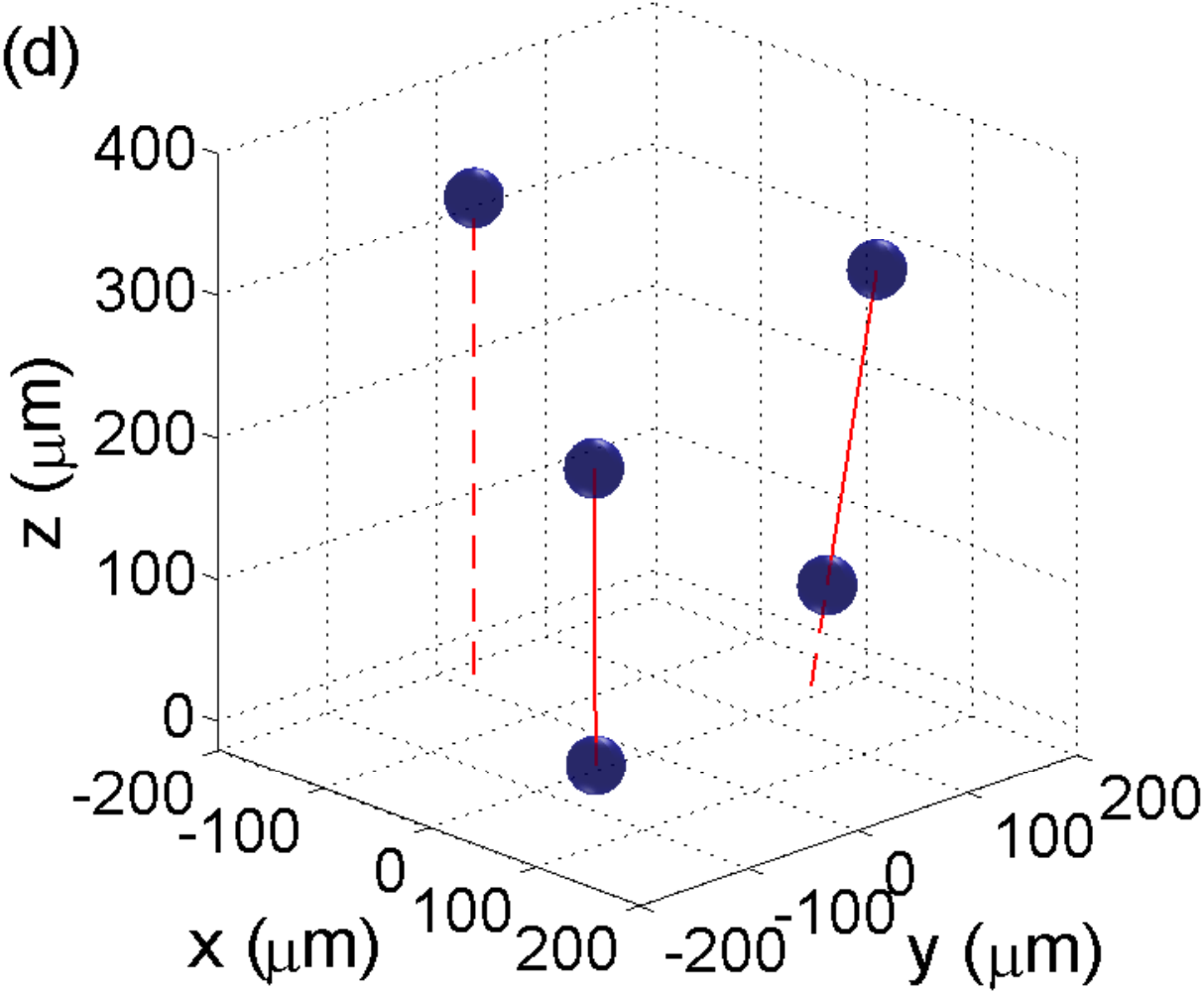}
}
\caption{Chirality of structures. (a) Side view of the 5-particle cluster shown in Fig. \ref{fig:asymmetricStruc} (a), and a cluster which is its mirror image (b). System parameters for (b) are rf power 1.83 W, pressure 150 mTorr. Reconstructions of these structures are shown in (c) and (d) where the blue dots represent the dust particles, the solid red lines indicate particles which are nearly vertically aligned, and the dashed line shows the projection of the unpaired particle on the $xy$-plane.}
\label{fig:helix}
\end{center}
\end{figure}

\begin{figure}[tbp]
\centering
\subfigure{
\includegraphics[width=0.21\textwidth]{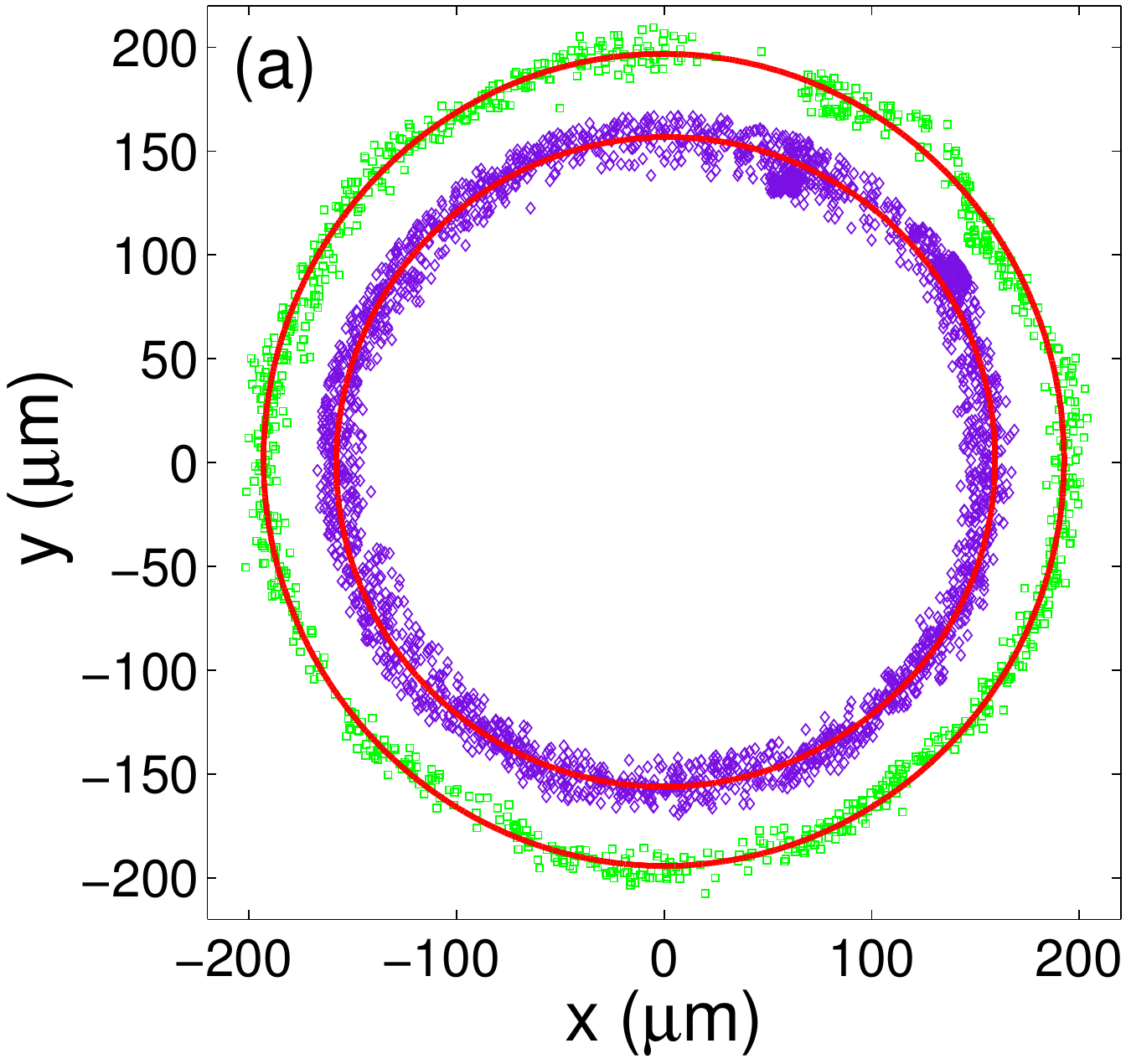} 
} 
\hspace{0.6pt}
\subfigure{
\includegraphics[width=0.21\textwidth]{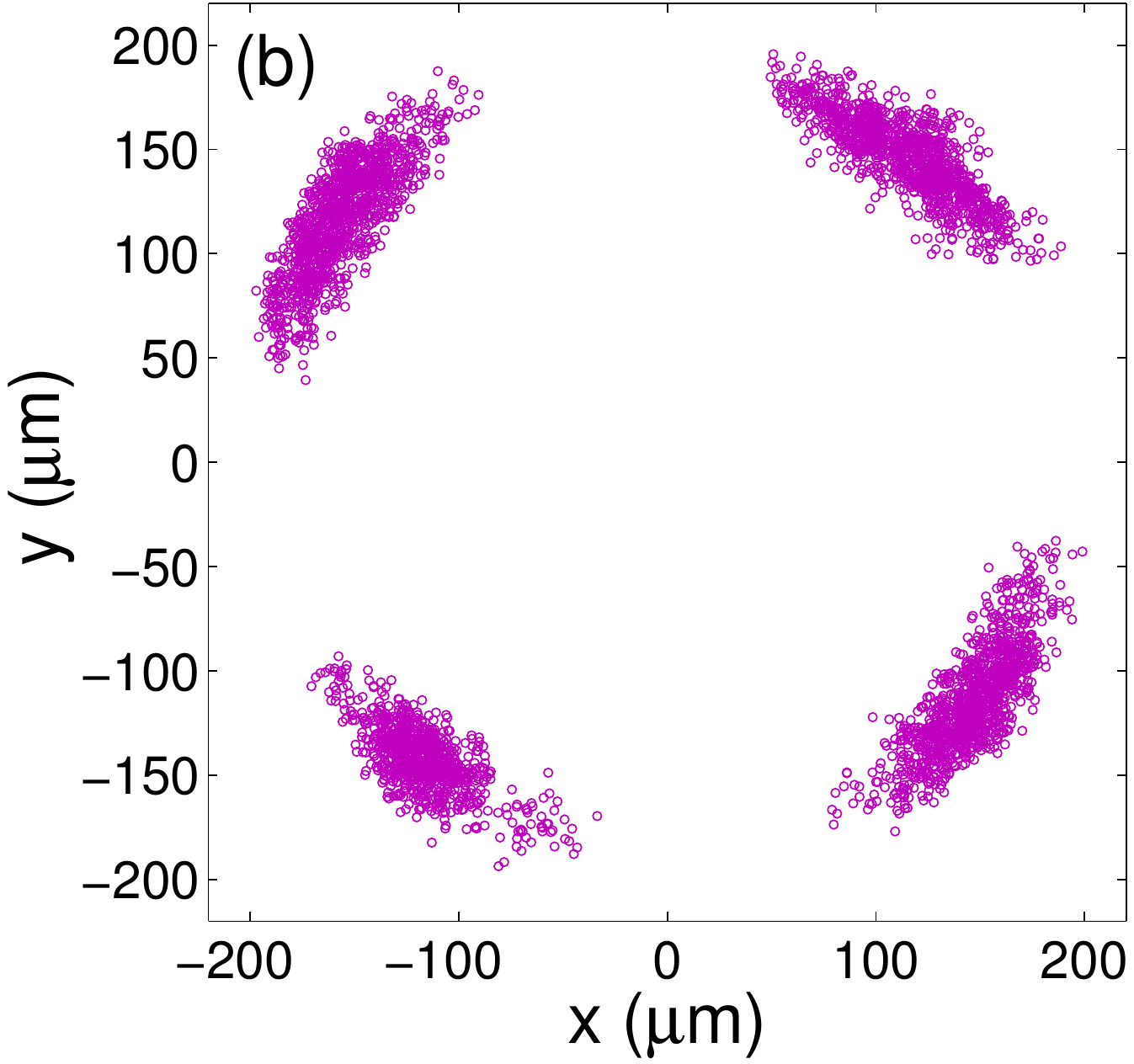} 
}
\subfigure{
\includegraphics[width=0.215\textwidth]{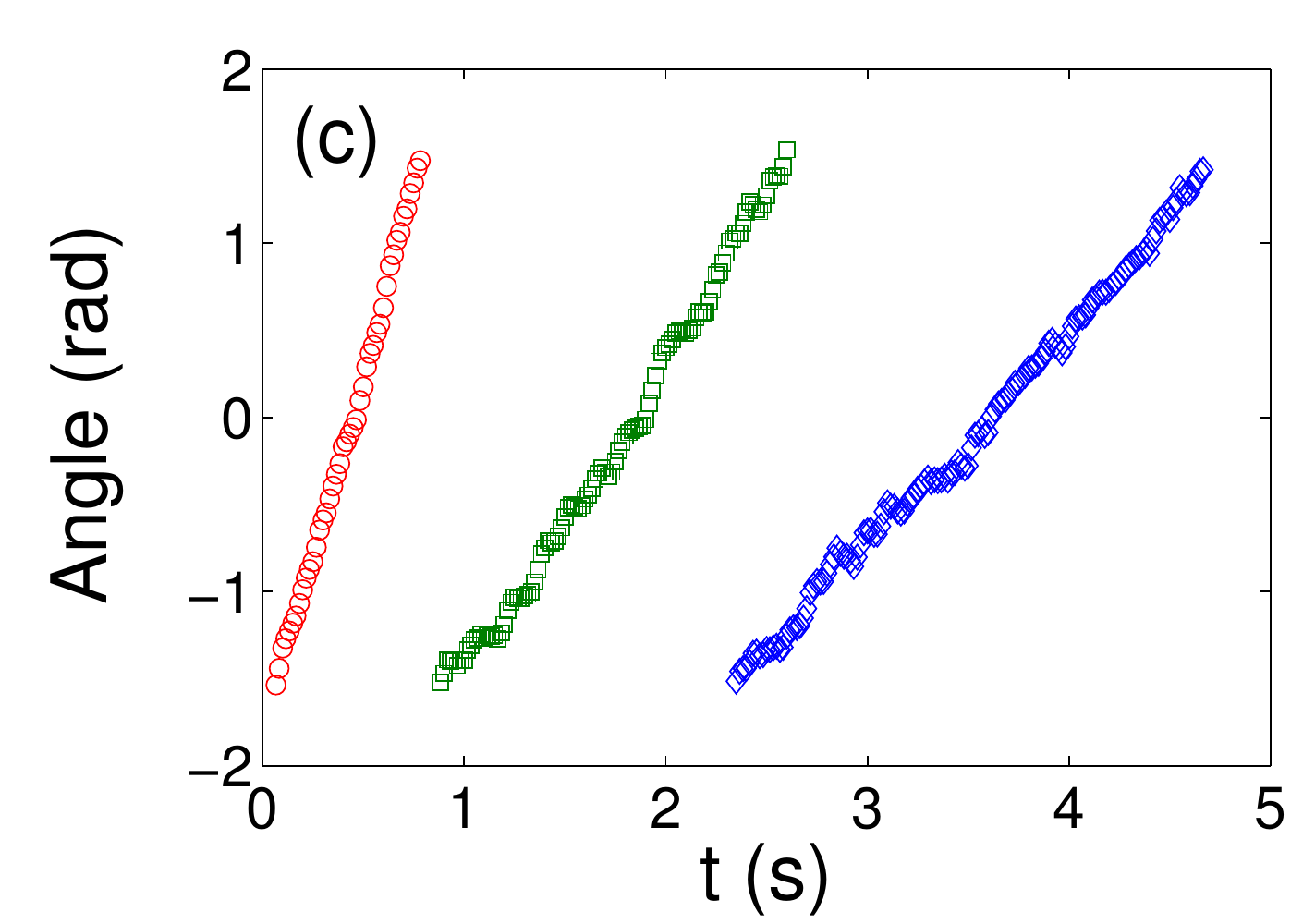}
}
\subfigure{
\includegraphics[width=0.215\textwidth]{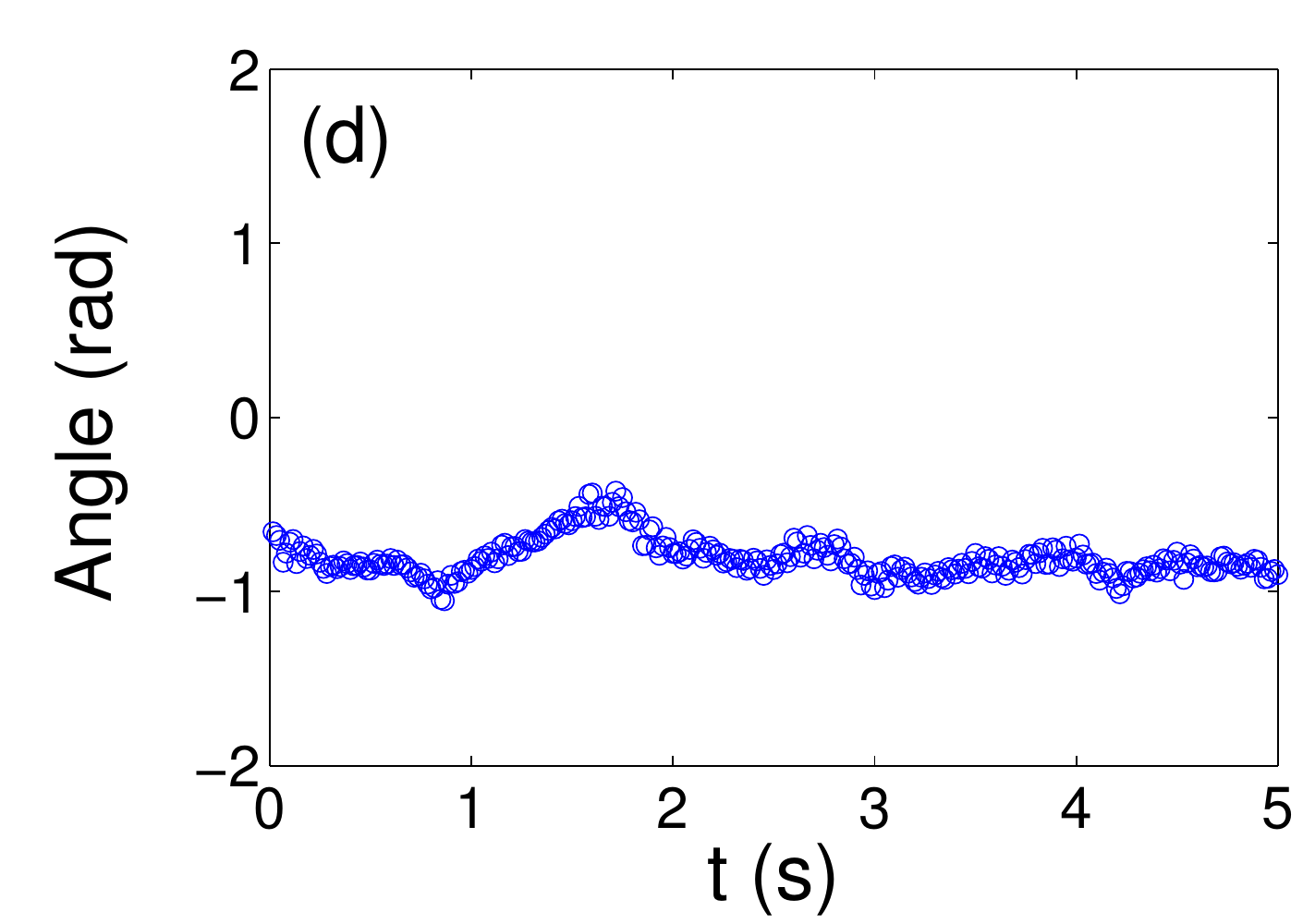}
}
\caption{Particle positions recorded over 20 seconds for (a) the asymmetric $N = 5$ cluster shown in Fig. 3 and (b) the symmetric $N = 8$ cluster shown in Fig. 2 (c). The points shown in (a) indicate the positions of the two different particles within the cluster. The lines indicate the best-fit ellipses to the circular motion. The rotation angle of a representative particle within each of the clusters, with respect to a fixed axis, is shown in (c) and (d) as a function of time. In (c), the rotation angle is shown for rf powers of 1.76 W (red circles), 1.83 W (green squares) and 1.89 W (blue diamonds).} 
\label{fig:trajec}
\end{figure}

\begin{table*}[tbp]
\setlength{\tabcolsep}{1.16em}
%\centering
\caption{Experimentally measured and theoretically calculated parameters for six asymmetric and two symmetric clusters. Calculated values for $M$, $q_w$, and $z_w$ are shown for $Q_d = 13000$e with decharging of downstream grains assumed to be $0.8Q_d$.}
\label{tab:torqueTable1}
\begin{tabular}{lcccccccccc} \\
\hline
\hline
          N  &Power    &P       &Bias      &$\omega$       &$h_{COM}$        &$\tau_{d}$           &$\lambda_{De}$     &$M$       &$q_w$       &$z_w$                \\
                           &(W)                 &(mTorr)           &(V)               &(s$^{-1}$)        &(mm)        &($\times 10^{-12}$ N$\cdot \mu$m)        &($\mu$m)         &       &(e)       &$(\lambda_{De})$            \\
\hline
          5       &1.76               &150                &-24.5         &-3.4       &7.57        &-5.06                  &336        &0.97          &2132                  &1.20                                              \\
          5       &1.83               &150                &-25.5         &1.9        &7.48         &3.20                   &330        &1.04           &2057                  &1.30                                                \\
          5       &1.89               &150                &-36.2         &1.1        &7.37         &1.79                    &324        &1.11           &1973                 &1.39                                              \\
 \\
          5        &1.92               &150               &-25.0        &1.5          &7.58        &2.10                    &322       &1.10            &1982                  &1.38                                             \\
          5        &1.92               &140               &-26.2        &-2.3        &8.04         &-3.32                   &333       &1.03            &2069                  &1.28                                                \\
          5        &1.92               &130               &-27.3        &-3.7        &8.44         &-4.79                   &345       &0.99             &2115                  &1.23                                               \\
 \\
         6        &2.17               &150               &-35.3        &-0.029    &6.78         &-0.0751              &300        &1.12            &1964                  &1.40                             \\
         8        &2.56               &150               &-41.2        &0.055      &6.61        &0.211                  &269        &2.34            &1194                  &2.80                                          \\
\hline 
\hline
\end{tabular}
\end{table*}

\begin{figure}[tbp]
%\centering
%\vspace{1pt}
\subfigure{
\includegraphics[width=0.354\textwidth]{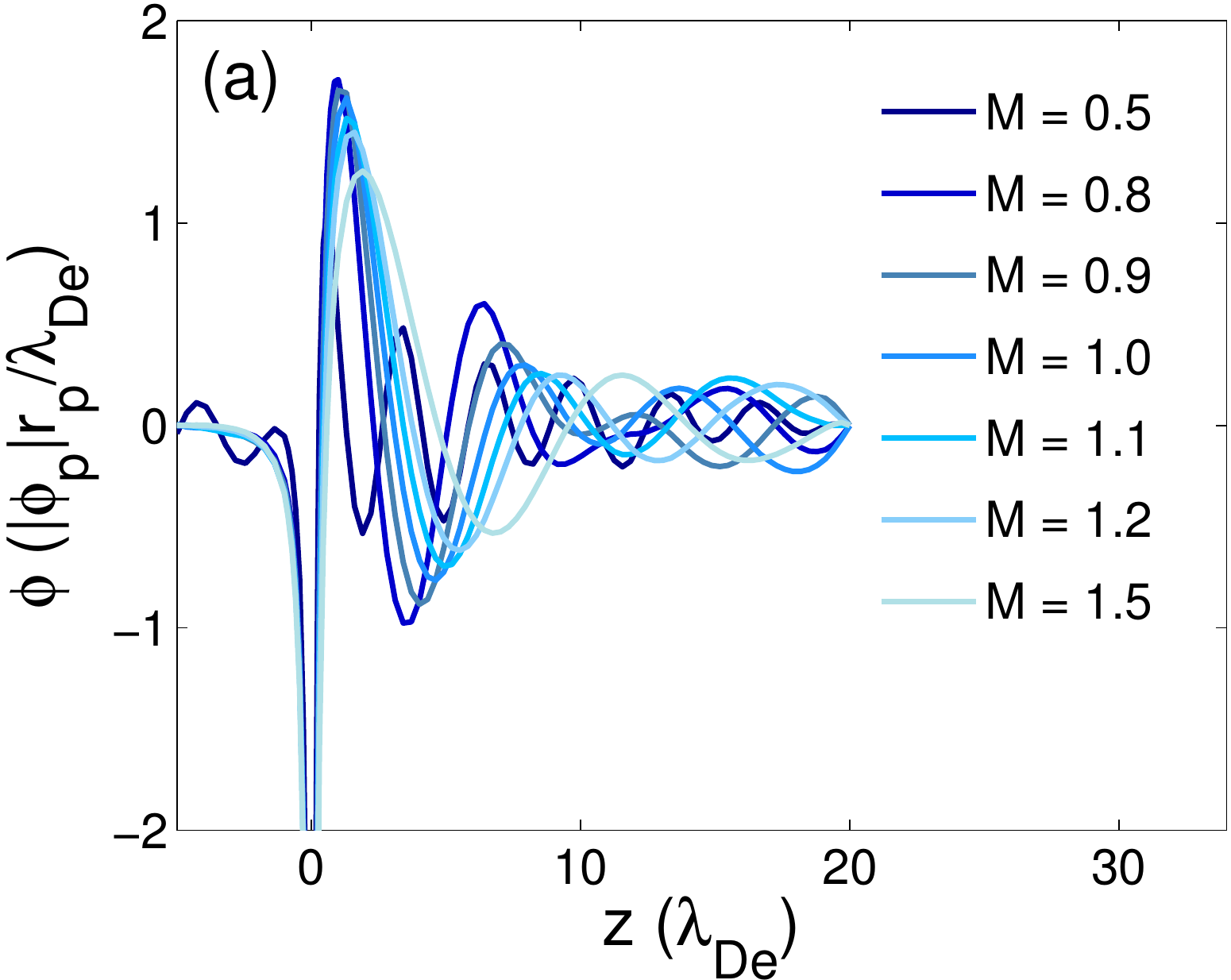}
}

\subfigure{
\includegraphics[width=0.358\textwidth]{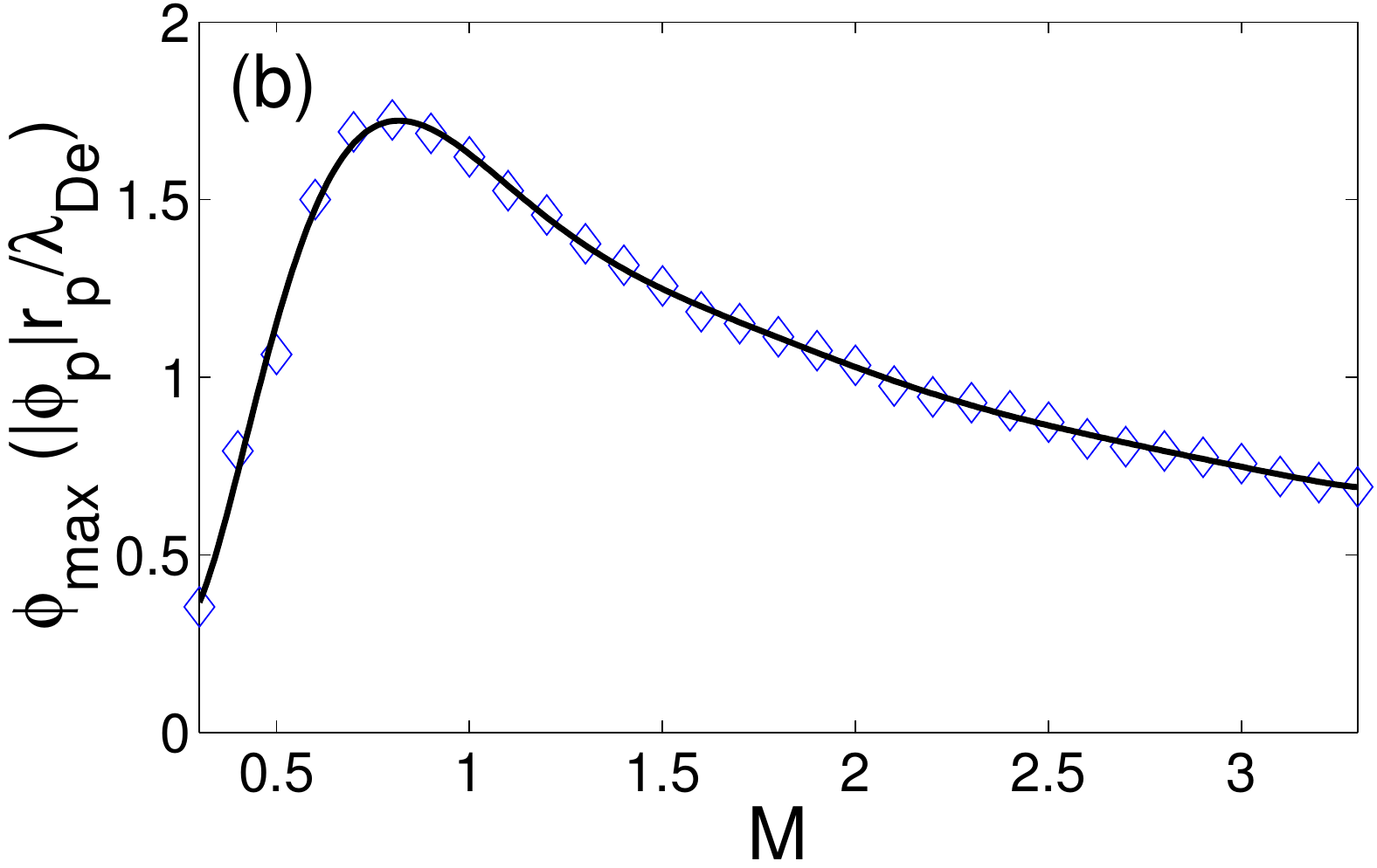}
}

\subfigure{
\includegraphics[width=0.358\textwidth]{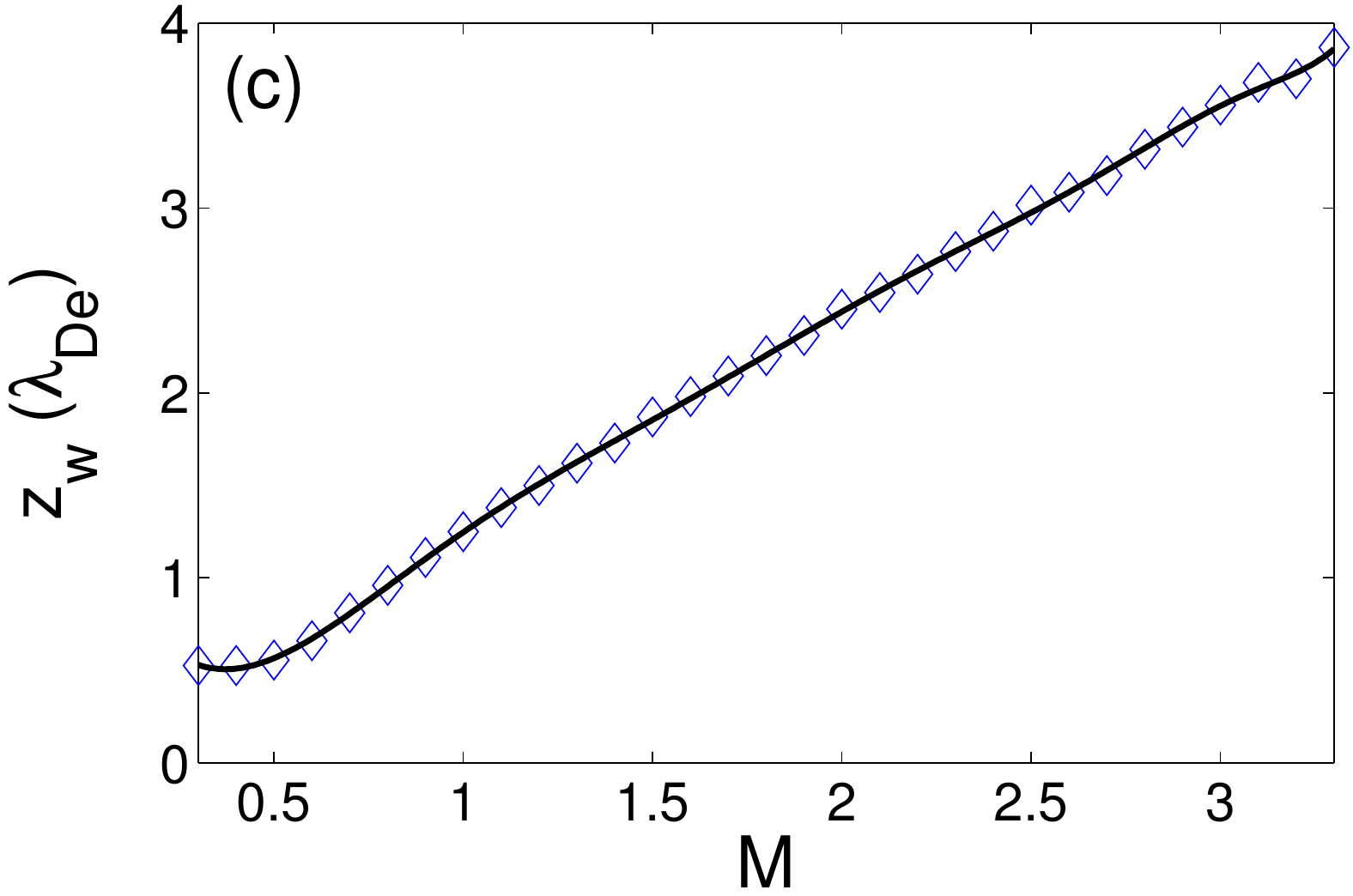}
}
\caption{(a) Normalized wake potential profile along the $x = y = 0$ axis for distinct ion drift velocities. Variation of the maximum wake potential is shown in (b) and its position with drift speed is given in (c). The lines in (b) and (c) provide a high order polynomial fit to the data.} 
\label{fig:wakeAnalysis}
\end{figure}

Once cluster symmetry was broken, spontaneous rotation was observed. An asymmetric $N = 5$ cluster is shown in Fig. \ref{fig:asymmetricStruc}. (A movie showing the complete rotation of this cluster is attached as Supplemental Material.) The direction of rotation of such asymmetric clusters can be either clockwise or counterclockwise, depending on cluster chirality. (See Fig. \ref{fig:helix} (a) and (b), for two five-particle clusters, along with their reconstructed 3D models as shown in Figs. \ref{fig:helix} (c) and (d).) The clusters shown in Fig. \ref{fig:helix} (a) and (b) rotated counterclockwise and clockwise, respectively, once formed. (Two movies are attached in the Supplemental Material to demonstrate this chirality-related rotation.) 

\begin{figure}[tbp]
\centering
\subfigure{
\includegraphics[width=0.37\textwidth]{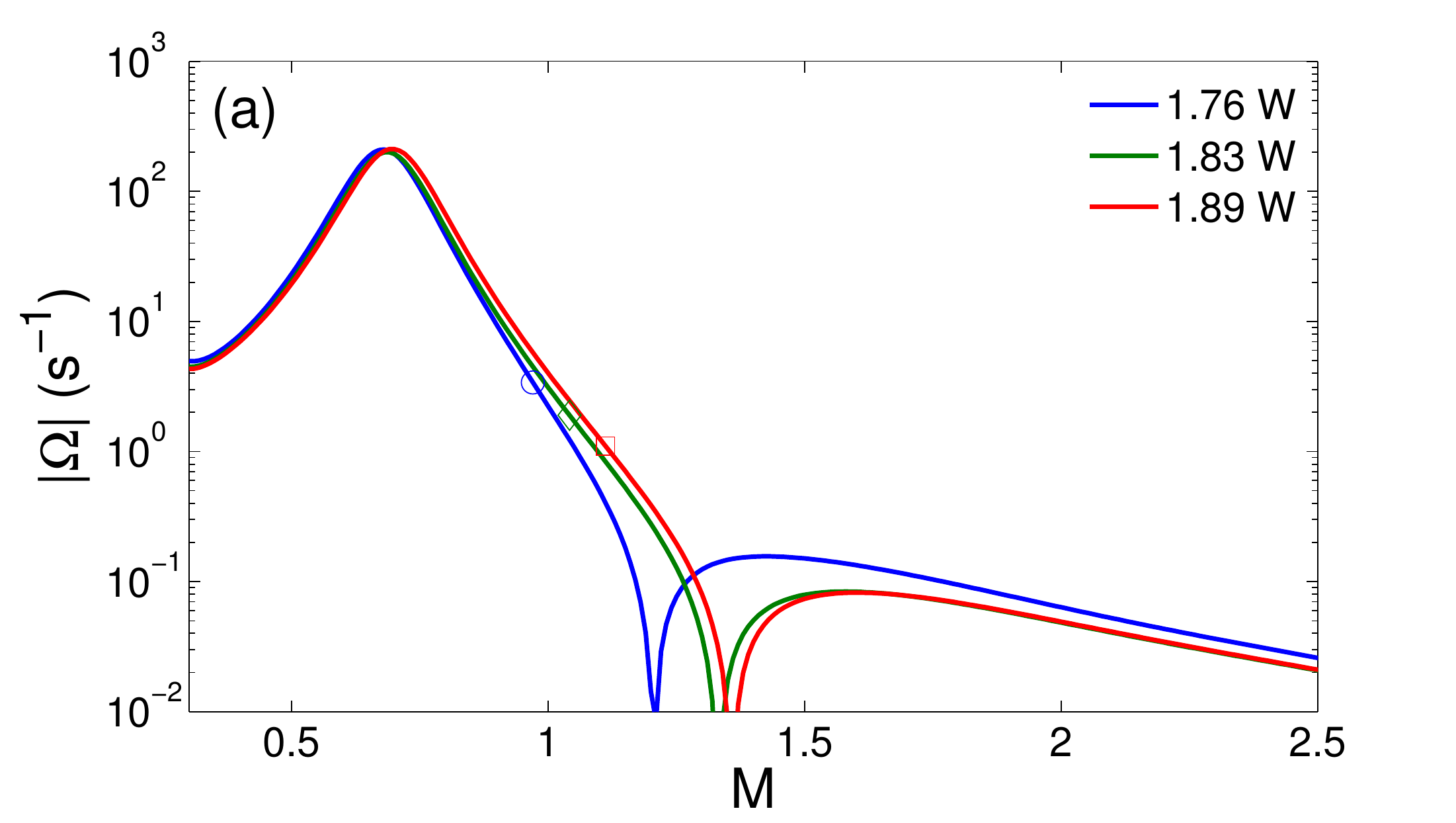}
}

\subfigure{
\includegraphics[width=0.37\textwidth]{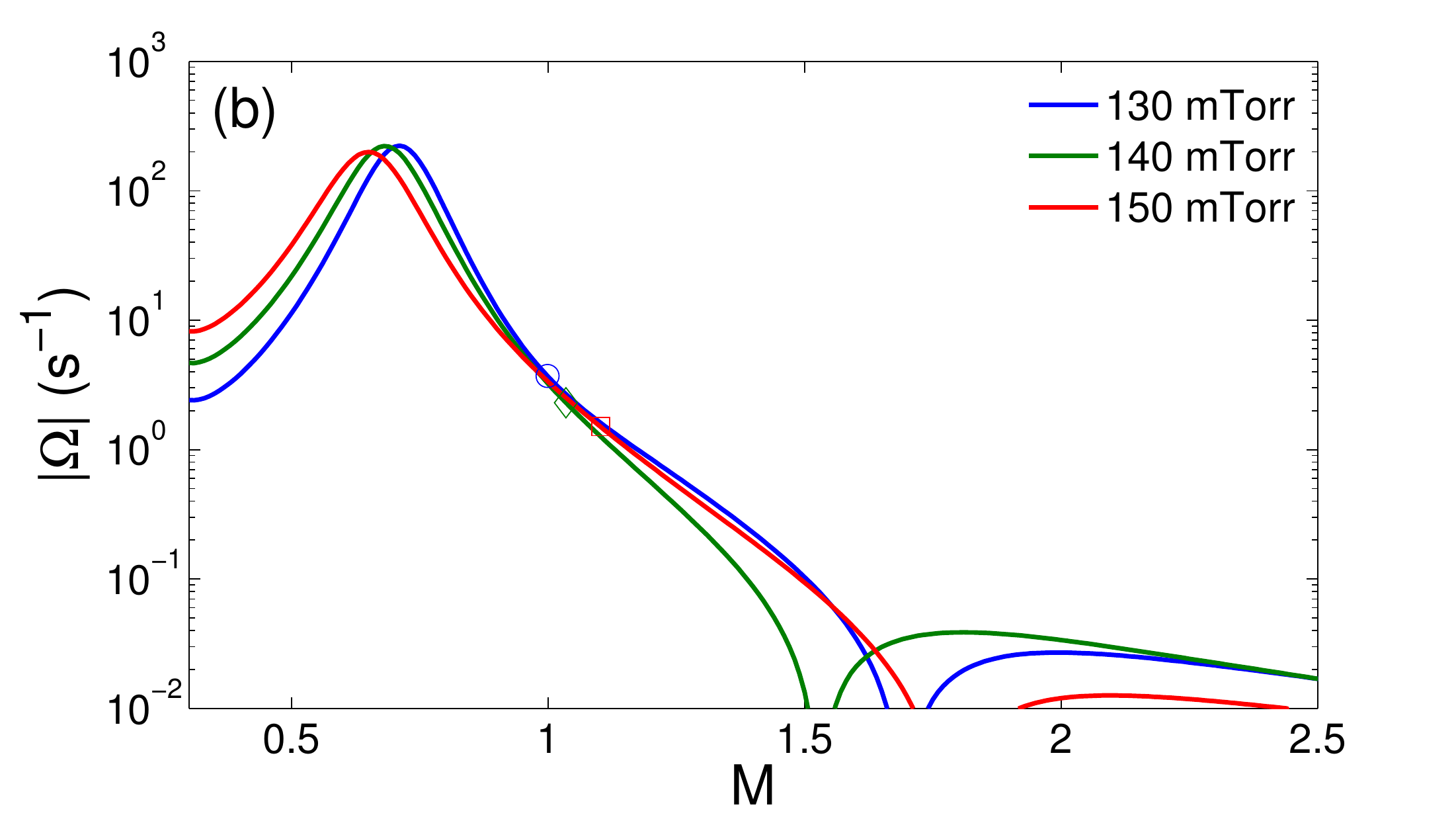}
}

\subfigure{
\includegraphics[width=0.37\textwidth]{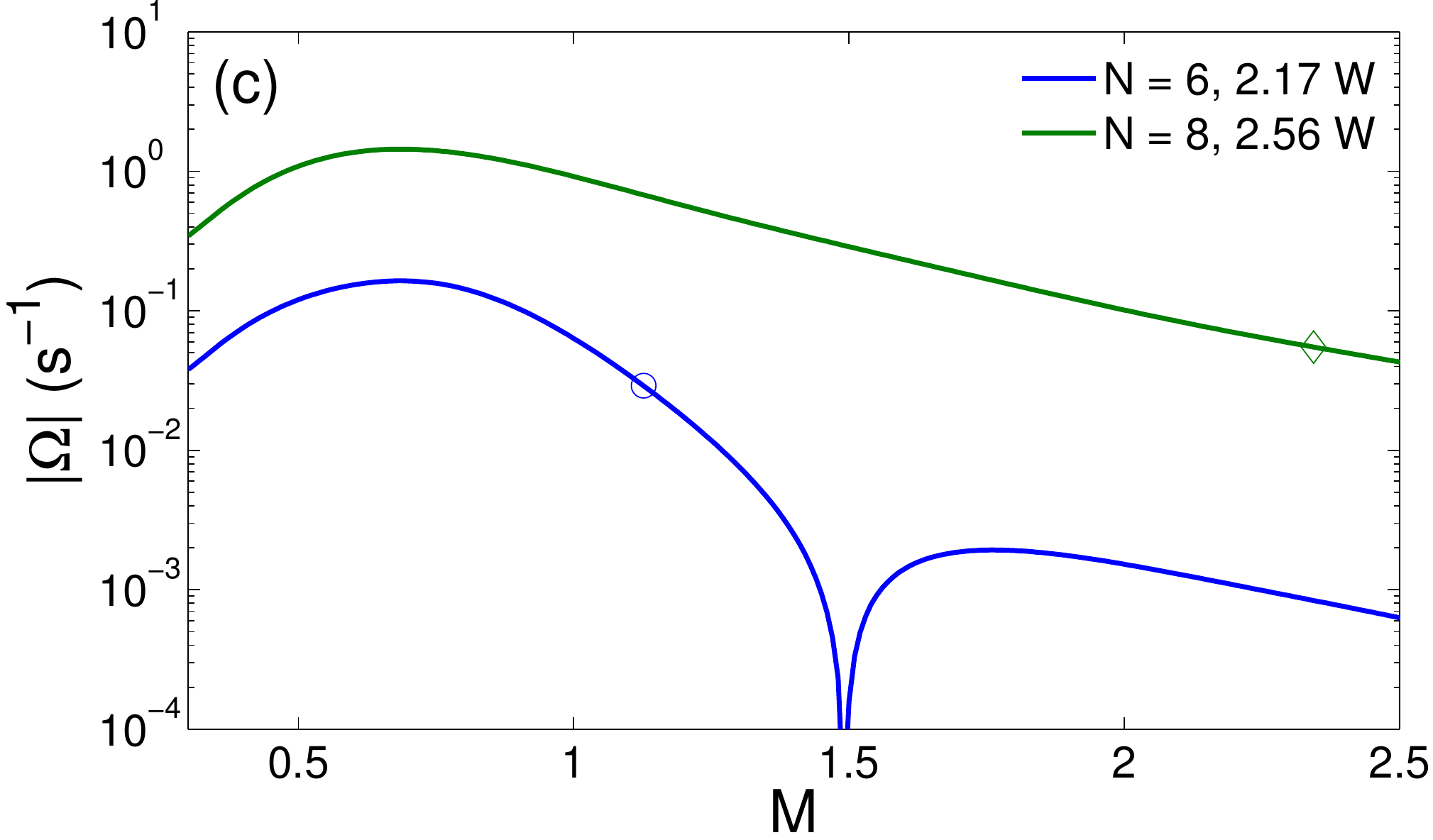}
}
\caption{Calculated rotation speeds $\Omega$ of asymmetric $N = 5$ clusters as a function of the ion drift velocity for a fixed pressure of 150 mTorr (a) and a fixed power of 1.92 W (b). Variation of $\Omega$ with drift speed for two symmetric structures (c) at a pressure of 150 mTorr. Here $Q_d = 13000$e and the downstream grains are assumed to have a charge $0.8Q_d$. Symbols indicate the drift velocity required to match the rotation speed observed experimentally.} 
\label{fig:omegaAnalysis}
\end{figure}

\begin{table}[tbp]
\setlength{\tabcolsep}{1.15em}
%\centering
\caption{Ion drift speed determined using varying values of $Q_d$ and the decharging factor of downstream grains under specific rf powers at 150 mTorr.}
\label{tab:machTable}
\begin{tabular}{l c c c c} \hline\hline
\multicolumn 1 {c}{Power} &
\multicolumn 1 {c}{Decharging} &
\multicolumn 3 {c}{$Q_d (e)$} \\ \cline{3-5}
 (W)              &$(Q_d)$           &12000  &13000     &14000 \\ \hline
1.76      &0.7      &0.92      &0.96         &0.97 \\ 
1.76             &0.8      &0.95      &0.97         &0.98 \\ 
1.76              &0.9      &0.96      &0.97         &0.98 \\ 
1.83      &0.7      &1.01      &1.03         &1.04 \\ 
1.83              &0.8      &1.02      &1.04         &1.05 \\ 
1.83             &0.9      &1.03     &1.05         &1.06 \\ 
1.89      &0.7      &1.08      &1.10         &1.11 \\ 
1.89            &0.8      &1.09      &1.11         &1.12 \\ 
1.89             &0.9      &1.10      &1.12         &1.13 \\ \hline\hline
\end{tabular}
\end{table}

Fig. \ref{fig:trajec} (a) and (b) illustrate representative particle trajectories in the horizontal plane (i.e., imaged by the top view camera) for the asymmetric cluster shown in Fig. \ref{fig:asymmetricStruc} and symmetric structure shown in Fig. \ref{fig:symmetricStruc} (c), respectively. Interestingly, the center of  rotation for the asymmetric cluster is not located at the projection of the cluster's COM in the horizontal plane, which is given by $\vec{r}_{com} =  \sum \vec{r}_i/N$, where $\vec{r}_i$ is the position of each dust particle with respect to the center of the box and height above the lower electrode and $N$ is the total number of particles comprising the cluster. Fig. \ref{fig:trajec} (c) and (d) show the rotational orientation of the clusters over time. As shown, the asymmetric cluster exhibits a uniform angular rotation speed, which increases as the power is decreased (Fig. 5(c)) while Fig. \ref{fig:trajec} (d) shows only a small change in orientation of the symmetric cluster, with a maximum rotation speed of 0.055 s$^{-1}$. The angular speed $\omega$ and  the height of the COM $h_{COM}$ of the clusters are summarized in Table \ref{tab:torqueTable1} for various experimental conditions. For fixed rf power, the angular speed of the cluster decreases with increasing pressure as shown.   

\section{Rotational mechanisms and discussion}

We propose that the spontaneous rotation observed for the small clusters described in this experiment is induced by the torque due to the ion wake field force when applied to the cluster once cluster symmetry is broken. At equilibrium, the net torque $\vec{\tau}_{net}$ that causes the cluster's rotation is balanced by the torque $\vec{\tau}_{d}$ due to the neutral drag force created by the cluster's uniform rotation, \begin{equation} \vec{\tau}_{net} = \vec{\tau}_{d}. \end{equation} The neutral drag torque $\vec{\tau}_{d}$ is given as \begin{equation} \vec{\tau}_{d} = \sum_{i=1}^{N'}\vec{r_j} \times \vec{F}_{dj}, \end{equation} where $\vec{F}_{dj}  = -m_d\beta\vec{v}_j$ is the neutral drag force applied on the $j$th particle, $\vec{r}_j$ is the position of the particle measured from the axis of rotation, $\vec{v}_j$ is the tangential velocity of the $j$th dust particle and $\beta$ is the Epstein drag coefficient defined as \begin{equation} \beta = \delta\frac{8}{\pi}\frac{P}{a\rho v_{th,n}}, \end{equation} where $\delta$ is a coefficient depicting the reflection of the neutral gas atoms from the surface of the dust, ($\delta = 1.26$ $\pm$ $0.13$ for MF particles in argon gas ~\cite{binliu}), $P$ is the gas pressure, $a$ the particle radius, $\rho$ is the particle's mass density and $v_{th,n} = \sqrt{8k_BT_n/\pi m_n}$ is the mean thermal velocity of the neutral gas. The temperature of the neutral gas is taken to be $T_n = 300$ K and the mass of the argon gas $m_n = 6.64 \times 10^{-26}$ kg. The neutral drag torque $\vec{\tau}_{d}$ is calculated and summarized in Table \ref{tab:torqueTable1} for four different clusters under different experimental conditions. 

The torque driving the rotation can be written as \begin{equation} \vec{\tau}_{net} = \sum_{j=1}^{N} \vec{r_j} \times \vec{F}_j, \end{equation} where $\vec{F}_j$ is the total force exerted on the $j$th particle excluding the neutral drag force, and is given by \begin{equation} \vec{F}_{j} = \vec{F}_{elecj} + \vec{F}_{ionj} + \vec{F}_{interj} \end{equation} with $\vec{F}_{elecj}$ being the electric field force,  $\vec{F}_{ionj}$ the ion wake field force, and $\vec{F}_{interj}$ the interparticle force from all other particles exerted on the $j$th particle. Inasmuch as $\vec{F}_{elecj} = \vec{\nabla} U_j$ is a conservative force, zero work is done by $\vec{F}_{elecj}$ moving a dust particle through a complete rotational trajectory, i.e. $\oint \vec{F}_{elecj} \cdot d\vec{r}_j = 0$. As such, $\vec{F}_{elecj}$ will not produce steady-state rotation for either symmetric or asymmetric structures since it cannot feed energy to the system ~\cite{flanagan09}. $\vec{F}_{interj}$ is an internal force between dust particles and can not contribute to rotation of the structures. This leaves the ion wake field force $\vec{F}_{ionj}$ as one possible contributor to the observed rotation. Thus, Eq. (4) can now be rewritten as \begin{equation} \vec{\tau}_{net} = \sum_{j=1}^{N}\vec{r_j} \times \vec{F}_{ionj}. \end{equation} 

In order to determine $\vec{F}_{ionj}$, a point charge model was employed to model the ion wake field ~\cite{qiao13, qiao14, vladimirov95, goree95, ishihara97, lampe00, ivlev03, kompaneets07, zhdanov09}. Assuming the ion wakefield acts as a positive point charge located beneath each dust particle, the ion wake field force experienced by the $j$th dust particle $\vec{F}_{ionj}$ is given by \begin{equation} \vec{F}_{ionj} = \sum_{k \neq j}^{N}\frac{Q_dq_w(\vec{R}_k - \vec{r}_j)}{4\pi\varepsilon_0|\vec{R}_k - \vec{r}_j|^3}, \end{equation} where $Q_d$, $q_w$ are the dust charge and wakefield point charge, $\vec{R}_k = \vec{r}_k - z_w\hat{z}$ is the location of the point charge located a distance $z_w$ beneath the $k$th particle, and $\varepsilon_0$ is the vacuum permittivity. Since the cluster's rotation axis is in the vertical direction, only the horizontal component of $\vec{F}_{ionj}$ contributes to the driving torque for the rotation. 

The location ($z_w$) and magnitude ($q_w$) of the wakefield point charge depends on the experimental conditions, since the power and pressure settings determine the particle charge and ion drift speed. Changes to the rf power also alter the electron and ion density, as well as change the electron temperature, which determines the bias on the lower electrode (establishing the ion drift velocity) and the dust surface charge. 

The electron Debye length under representative experimental conditions was estimated based on the results presented in Ref. ~\cite{kong14, nosenko14} (see Table \ref{tab:torqueTable1}). The charge on a dust grain within the sheath of a rf discharge is generally on the order of 1000e per $\mu$m diameter. Using the result from previous experiments under similar experimental conditions the dust charge was assumed to be $\sim$12700e ~\cite{kong14}. However, we analyzed the motion assuming $Q_d = 12000$e, 13000e, and 14000e to determine the extent to which the dust charge influences the rotation rate. Additionally, theoretical and experimental results have shown that downstream dust grains are decharged relative to the upstream grains ~\cite{block15}. Accordingly, a charge of 0.7, 0.8, and 0.9$Q_d$ was assumed for the lower grains in a cluster. (See Table \ref{tab:machTable}.) 

Estimates for the point charge and its location downstream from a particle were obtained employing the COPTIC (Cartesian mesh, oblique boundary, particles and thermals in cell) code developed by Hutchinson ~\cite{hutch11pop, hutch11prl, hutch12pre, hutch13pop, hutch13ppcf}. In this simulation, grains are represented as point charges immersed in a collisionless plasma using uniform external drifting-Maxwellian ion distributions with $T_i/T_e = 0.01$. Calculations are performed on a $44 \times 44 \times 96$ cell grid with nonuniform mesh spacing over a cubical domain of $ 10 \times 10 \times 25$ Debye lengths, where the ions are flowing along the $\hat{z}$-direction with a drift velocity $v_d$ expressed as a Mach number $M = v_d/c_s$ where $c_s = \sqrt{T_e/m_i}$ is the cold-ion sound speed. The code is run with the point charge representing a dust particle located at position $(0,0,0)$. The analytical part of this point charge extends to radius $r_p = 0.1 \lambda_{De}$. At this distance, the floating potential $\phi_p = -0.25T_e/e$ with $T_e = 2.585$ eV. Distinct drift velocities ranging from 0.1 to 3.3 were used in the COPTIC program to determine the maximum value of the wake potential and its location. 

Fig. \ref{fig:wakeAnalysis} (a) shows the wake potential profile along the axial direction normalized to the dust grain potential as a function of the ion drift velocity $M$. As can be seen, the maximum wake potential $\phi_{max}$ achieves its peak value for $M = 0.8$ (Fig. \ref{fig:wakeAnalysis} (b)), with its position $z_w$ shifting further away from the dust grain for increasing drift velocity (see Fig. \ref{fig:wakeAnalysis} (c)). The magnitude of the wakefield point charge $q_w$ can be calculated as $q_w \approx Q_d\phi_{max}/|\phi_p|$, where $\phi_p$ is the value of the dust surface potential. 

The theoretical rotation speed $\Omega$ can now be calculated based on $q_w$ and $z_w$ where using the COPTIC model to determine the ion flow velocity which best matches the experimental results. Assuming $\lambda_{De}$ for the experimental conditions shown in Table \ref{tab:torqueTable1}, the magnitude and location of the wake point charge over the range of ion drift velocities were fit employing a higher order polynomial, and then used to calculate the torque on each of the clusters listed in Table \ref{tab:torqueTable1}. This torque was then equated to the neutral drag torque (Eq. 2) to determine the values of $q_w$ and $z_w$ needed to balance the torques, allowing an estimate of the ion drift speed to be obtained. As shown in Table \ref{tab:machTable}, the ion drift speed found using all possible values of $Q_d$ varies by less than $1.8\%$. The results of these calculations assuming an upstream dust charge $Q_d = 13000$e and all estimates of decharging for the downstream particle are shown in Fig. \ref{fig:omegaAnalysis} and Table \ref{tab:torqueTable1}.

As shown in Fig. \ref{fig:omegaAnalysis}, there are always two values of $M$ which produce a rotational speed matching the experimentally measured value, one for $M < 0.7$ and one for $M > 0.7$. As observed in this experiment, the levitation height of the cluster decreases as the power is increased, as does the rotation rate. As the ion drift velocity $M$ increases closer to the lower electrode ~\cite{douglass11} the trend for decreasing $\Omega$ with increasing power (and thus increasing $M$) points to ion drift velocities $> 0.7$, as shown in Fig. \ref{fig:omegaAnalysis} (a). At fixed power, reducing the pressure causes a cluster's rotation speed to increase while the height of its COM increases. Thus, the Mach number should decrease with decreasing pressure, as is seen in Fig. \ref{fig:omegaAnalysis} (b). 

Finally, as shown in Fig. \ref{fig:omegaAnalysis} (c), the rotation rates for symmetric clusters are very small over a wide range of ion drift velocities. Calculated ion drift speeds are consistent with the expected increase in these values, given the power range explored in the experiment. 

According to the trends shown in Fig. \ref{fig:omegaAnalysis} (a) and (b), as the power and pressure are increased further, the rotation speeds of the clusters should be reduced to almost negligible amounts. However, this was not observed experimentally since when the power or pressure exceeded certain critical values, the structure of the cluster changed. Thus asymmetric structures were always observed to have rotation rates on the order of 1-10 rad/s. 

\vspace{7pt}

\section{Conclusions}
Clusters of a small number of dust particles were produced within a glass box placed on the lower electrode of a GEC rf cell.
 
Self-excited rotation was observed for asymmetric structures with a uniform rotation speed, whereas no appreciable rotation was produced for symmetric structures. The asymmetric clusters were found to rotate about a vertical axis not passing through the center of mass. 
 
It was proposed that the spontaneous rotation for the small asymmetric dust clusters observed is probably induced by the net torque applied on the cluster due to the ion wake force. It was shown that symmetric clusters experience a very small net torque, and do not rotate. The rotation direction of the asymmetric cluster was determined by the conformational chirality of the specific structure, causing the cluster to spin either clockwise or counterclockwise. 
 
The torque induced by the ion wake force was calculated employing the ion wakefield point charge model, where the magnitude and location of the wakefield point charge was determined using the COPTIC code. Balancing the opposing torques induced by gas drag and the ion wake field allows the ion flow to be estimated within the glass box, which was found to be $\sim$1.0 $M$ for rf power $1.7 < P < 2.0$ W. This result is consistent with the values generally assumed for these experimental conditions ~\cite{hutch11prl}. These results are in rough agreement wiht Nosenko $et$ $al.$ ~\cite{nosenko12} who also found rotating particle pairs which they ascribed to the interaction with the ion wake field. Using the theory of Lampe $et$ $al.$ ~\cite{lampe12}, they estimated the Mach number of the ion flow to be 2.26 in the plasma sheath of an rf discharge at 157 mTorr and 5 W. Higher Mach numbers were also suggested by the results of this experiment in analyzing the small rotations observed for symmetric clusters formed at higher rf power.  

\section*{Acknowledgment}
Support from NSF/DOE Grant No. 1414523 and NSF/NASA Grant No. 1740203 is gratefully acknowledged.

\end{document}